\documentclass[prl,aps,preprint,superscriptaddress,showpacs]{revtex4-1}
\usepackage{graphicx}
\usepackage{amsmath, amsthm, amssymb}
\usepackage{bm}
\begin{document}

\title{Stability Peninsulas on the Neutron Drip Line}


\author{V.~N.~Tarasov} 
\affiliation{NSC, Kharkov Institute of Physics and Technology, Ukraine}
\author{K.~A.~Gridnev} 
\affiliation{Saint Petersburg State University, Russia}
\author{D.~K.~Gridnev}
\affiliation{Saint Petersburg State University, Russia}
\affiliation{FIAS, Goethe University Frankfurt, Germany}
\author{D.~V.~Tarasov} 
\affiliation{NSC, Kharkov Institute of Physics and Technology, Ukraine}
\author{S. Schramm} 
\affiliation{FIAS, Goethe University Frankfurt, Germany}
\author{X. Vi{\~n}as} 
\affiliation{University of Barcelona, Spain}
\author{Walter Greiner} 
\affiliation{FIAS, Goethe University Frankfurt, Germany}

\begin{abstract}
Using HF+BCS method with Skyrme forces we analyze the neutron drip line. It is shown that around magic and new magic numbers the drip line 
may form stability peninsulas. It is shown the location of these peninsulas does not depend on the choice of Skyrme forces. 
It is found that the size of the peninsulas is sensitive to the choice of Skyrme forces and the most extended peninsulas appear with the SkI2 set. 
\end{abstract}

\maketitle

One of the major problems in nuclear physics having also interdisciplinary
importance is the positioning of the neutron drip line. The microscopic approaches to the study of neutron rich nuclei include the Hartree-Fock-Bogoliubov (HFB) and Hartree-Fock (HF) methods using effective forces, 
see f. e.  \cite{2,4,5} or relativistic meanfield (RMF) theory \cite{6}. 
The standard theoretical
procedure in
locating the neutron drip line is to take a stable nucleus and load it with
neutrons until it saturates, that is adding extra neutrons makes the isotope undergo the neutron decay thereby releasing
these extra neutrons. This method, however, implies a simple structure
of the drip line, namely, that any straight line on the nuclear chart, which
corresponds to a fixed number of protons, crosses the neutron drip line
only once. Yet it might happen that the drip line has a more complicated
structure. In the vicinity of magic numbers or new magic numbers the
following scenario can take place. At some point $(N, Z)$ nuclei loose their stability
but then after gaining more neutrons their stability becomes restored. This
leads
to the formation of stability peninsulas on the nuclear chart.

\begin{figure}[htp]
\includegraphics[totalheight=0.7\textheight]{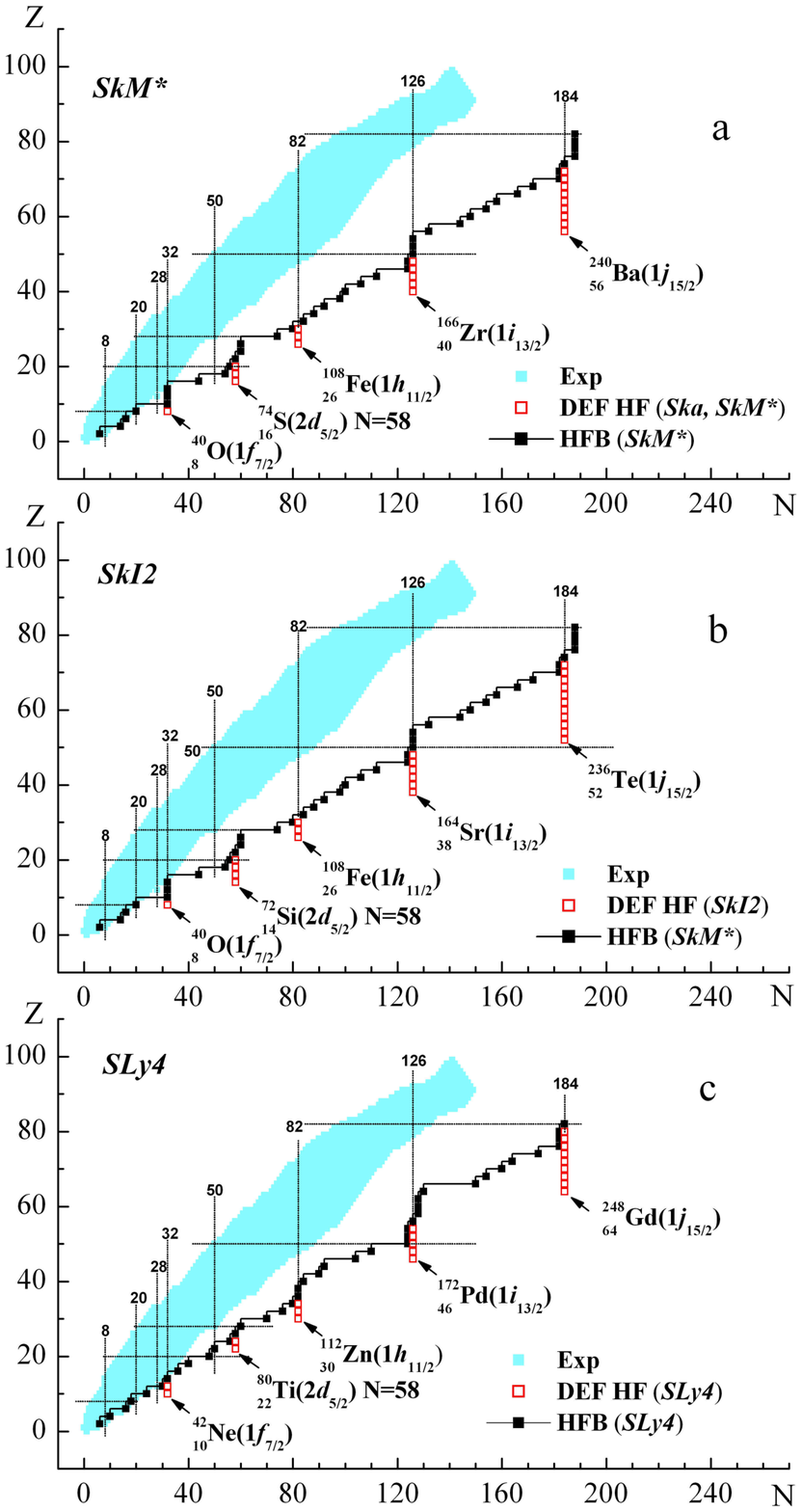}
\caption{(Color online). The nuclear chart and the 1n drip line. Filled blue area shows experimentally
known nulei. 
The solid line going through the dark squares is the 1n drip line in HFB
calculations \cite{4}. 
Red empty squares a nuclei stable against 1 neutron emission according to DEF HF
calculations with 
SkM* (a), SkI2 (b) and SLy4 (c) forces. The dotted line goes corresponds to 
magic numbers $N = 8, 20, 28, 50, 82, 126, 184$ and to the new magic number
N=32.}
\end{figure}

This scenario of formation of stability peninsulas has been analyzed by us in
Refs.~ \cite{9,11,12,13,14,15,16,16.1} for the isotopes 
of  O, Ar, Ni, Zr, Kr, Pb, Rn and other elements. The calculations were performed
using the HF+BCS approach with Skyrme forces accounting for 
deformations (DEF HF approach). In \cite{12,13,14,15,16} we have analyzed the mechanism, which
leads to the stability restoration. 
It turns out that nuclei lying close to such peninsulas possess low--lying
quasistable one--particle states. By adding neutrons one makes these levels 
dive into the discrete spectrum, thus stabilizing the isotope. The stability restoration by additional neutrons was also observed in \cite{16.2}, where 
the authors use the D1S Gogny force.

The discussion of the DEF HF method one finds in \cite{11,17,18,19}. The DEF HF
calculations are performed using the deformed harmonic oscillator basis,
where the basis parameters are optimized on each iteration, for details see
\cite{11}. The optimization procedure consists in choosing optimal oscillator
frequencies, which minimize the total energy within a given basis. Let us
mention that the optimization procedure facilitates the calculation and,
more important, corrects the basis functions for the spatially extended
density distributions near the drip line \cite{11}. 
The pairing constant is set to $G=(19.5/A)[1\pm 0.51(N-Z)/A]$, where the plus and minus signs refer to protons and neutrons respectively. In DEF HF
calculations we include only bound
one particle states. In spite of ignoring the continuum states this method
still provides a good agreement with the HFB, see
\cite{12,13,14,15,16}. 
Since all nuclei, which lie on
the stability peninsulas are spherical we also use a spherical code (SPH HF) \cite{19.1},
which solves the HF equations directly rather than using a particular basis. In
the BCS scheme of the spherical code we implement the inclusion of localized quasibound
continuum states, which are confined under the centrifugal barrier. 
The size of the box used to discretize the continuum is set to 50 fm. The quasi bound state is considered localized if the probability to find the particle 
beyond the radius corresponding to the maximum of centrifugal barrier is less than 50\%. 

We used the following sets of Skyrme forces Ska, SkM*, SLy4, and SkI2, see
\cite{20,21,22,23} 
and references therein (below under Skyrme forces we shall mean only these sets of forces). Stability peninsulas originate for all Skyrme forces,
which produce low--lying 
quasibound one--particle states with high angular momentum (which are
responsible for a high centrifugal barrier confining the particles). 

\begin{figure}[htp]
\includegraphics[totalheight=0.25\textheight]{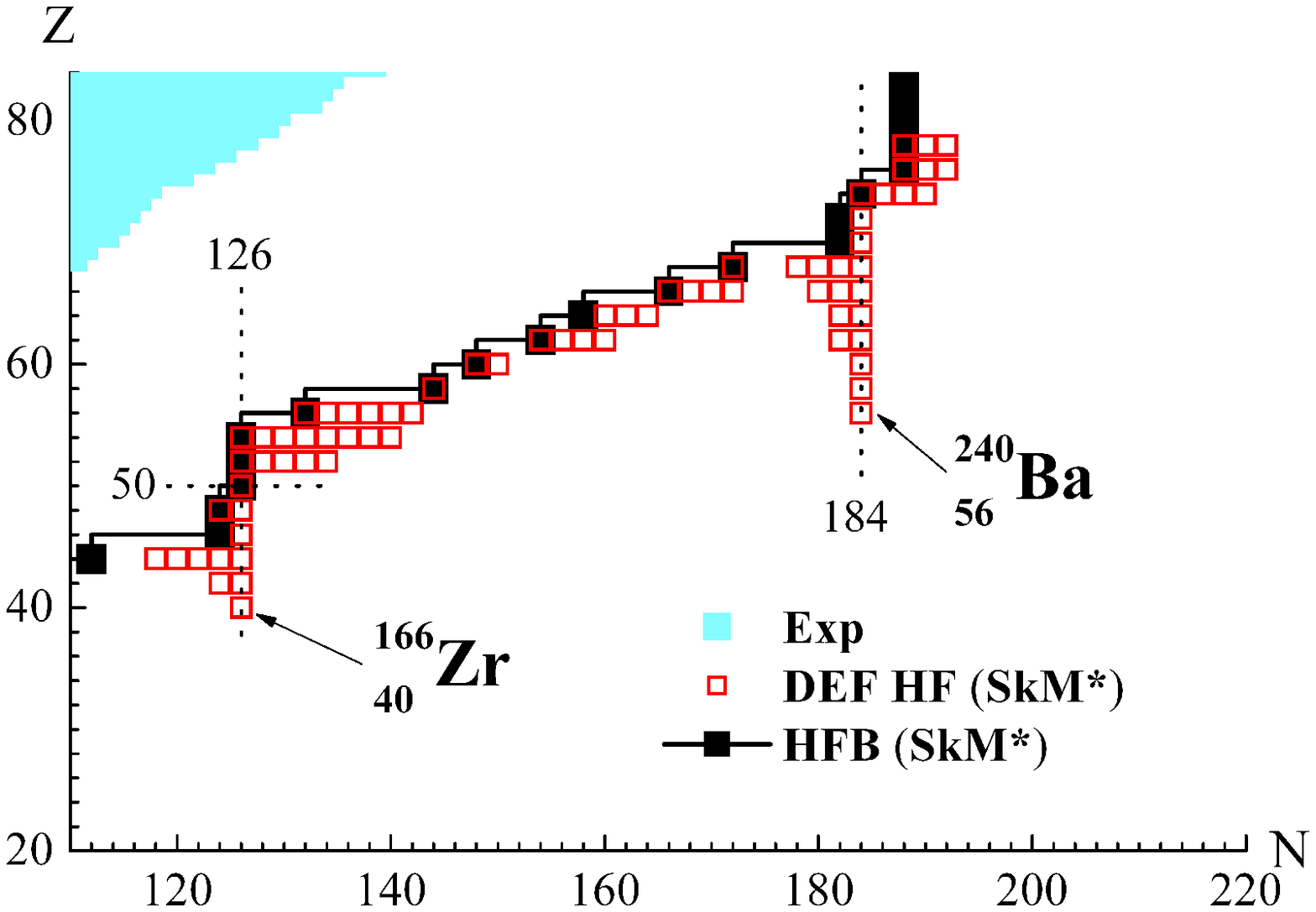}
\caption{(Color online). A detailed fragment of Fig.~1. The notations are that of Fig.~1.}
\end{figure}

By definition the one or two neutron separation energies on these lines would
be very small, sometimes within 0.1 Mev. Unfortunately, various Skyrme
forces do not agree with each other over the whole nuclear chart to that
degree of accuracy. So it would be naive to expect that different Skyrme
forces would all predict a unique neutron drip line. However, all Skyrme forces
predict the 
same magic and new magic numbers. The observed stability peninsulas with
respect to one neutron emission
are shown in Fig.~1 for different Skyrme forces. It is easy to see from
Fig.~1 that the formation of peninsulas on the neutron drip line happens
at the same N values for all forces but peninsula edges have different Z
values.

It should be stressed that nuclei forming
stability peninsulas are spectrally bound in the sense that there exists
a well-defined ground state wave function, which minimizes the energy
functional for such nuclei. At this point they become well-defined compact
objects and the question about their lifetime becomes correctly formulated.
And though for some nuclei it may be energetically favorable to get rid of
two or more neutrons, a large centrifugal barrier of the last filled levels may
serve as an indication that this lifetime would be large. Let us also mention that all spectrally stable nuclei in our
calculations appear as such in grid calculations based on the code from \cite{23.05}. 

We found that Ska and SkM* forces occupy an intermediate position
between SkI2 and SLy4, in the sense that more elements are stable with SkI2
and less with SLy4. The forces SkI2 are the most optimistic, at the same time
the formation of stability peninsulas for SLy4 is rather rare. The difference
between two of these forces is depicted in Fig.~1.

In Fig.~1 one finds the comparison with benchmark HFB calculations \cite{4}.
The position of 1n drip line in \cite{4} is obtained from the condition
$\lambda_n = 0$. 
In our method we use the Koopmans theorem to approximate one neutron separation
energies $S_n$ and define the drip line by condition $S_n = 0$, where even
small 
positive $S_n$ indicates the stability against the one--neutron emission. Note
that within HF and HFB methods one neutron separation energies can be
calculated 
only using certain approximations \cite{16.2,31} and 1n drip lines determined from the conditions
$\lambda_n = 0$ and $S_n = 0$ may not necessarily coincide \cite{16.2}.

1n drip line in Fig.~1 for various Skyrme forces has typical bend points around
known magic numbers  $N=82$ (SLy4), $N = 126$ (SkM*, SkI2, SLy4),
$N =184$ (SkM*, SkI2, SLy4) and also for $N=32$ (SkM*, SkI2) and $N= 58$ (SkM*,
SkI2, SLy4). The bend points indicate 
the stability enhancement around these $N$--values. It is worth noting that the
stability peninsulas are formed for various Skyrme forces 
around the same neutron numbers. As we have already mentioned the stability
restoration results from low--lying quasibound states, which immerse 
into the bound spectrum for higher $N$ \cite{12,13,14,15,16}. 

Below we list the stable isotopes that form stability peninsulas and the responsible subshells for the Skyrme
forces SkM* and Ska.  $1f_{7/2}$ -- $^{40}$O;  $2d_{5/2}$ -- $^{76}$Ar, $^{74}$S; 
$1h_{11/2}$ -- $^{110}$Ni, $^{108}$Fe;  $1i_{13/2}$ -- $^{174}$Cd, $^{172}$Pd,
$^{170}$Ru, $^{168}$Mo, $^{166}$Zr;  $1j_{15/2} $ --  $^{256}$Hf, $^{254}$Yb,
$^{252}$Er, $^{250}$Dy, $^{248}$Gd, $^{246}$Sm, $^{244}$Nd, $^{242}$Ce,
$^{240}$Ba. From Fig.~1 it can be seen that 
the stability peninsulas with SkI2 forces are by one or two 
$Z$ longer than those formed with SkM*. For $N=184$ the last stable
isotope with SkI2 forces is $^{236}$Te having $Z=52$, which is close to the
magic $Z=50$. The neutron to proton ratio in $^{236}$Te reaches $N/Z=3.54$. 
In the case of $^{40}$O one has $N/Z=4$! For SLy4 forces the whole
nuetron drip line becomes shifted 
in the positive $Z$ direction and the edges
of stability peninsulas are formed by the isotopes 
$^{42}$Ne; $^{80}$Ti, $^{112}$Zn, $^{172}$Pd, $^{248}$Gd. 
\vspace{.3cm}
\begin{figure}[htp]
\includegraphics[totalheight=0.45\textheight]{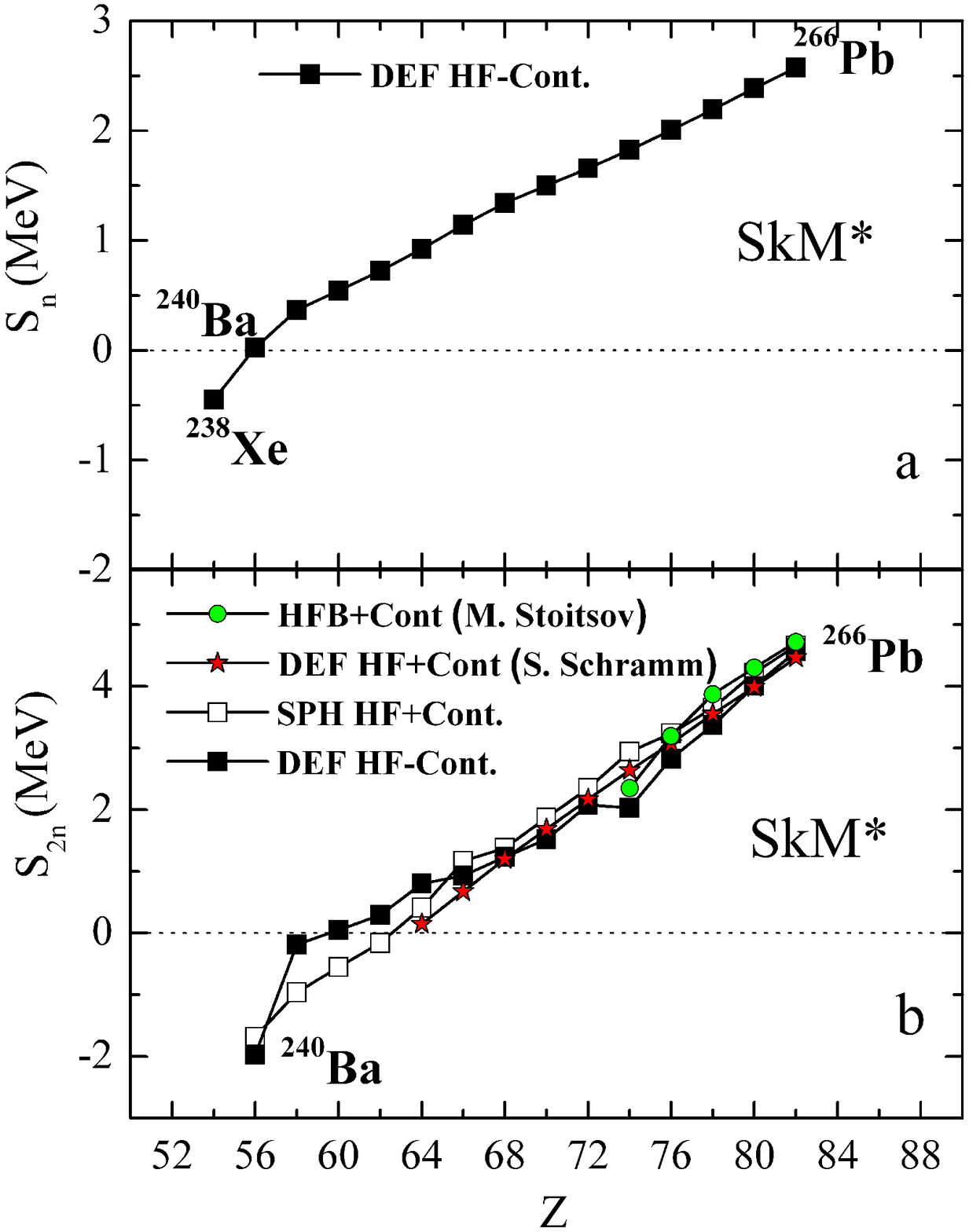}
\caption{(Color online). One and two neutron separation energies for the series of isotones
corresponding to $N=184$ for SkM* forces. (a) One neutron separation energies in
DEF HF calculations without continuum states in the BCS scheme. 
(b) two--neutron separation energies. Dark squares are DEF HF calculations (without
continuum in the BCS). Light squares are SPH HF calculations with continuum states.
Red stars are DEF HF in grid calculations with continuum states. Green circles 
are HFB calculations \cite{4}.}
\end{figure}

For all Skyrme forces the nuclei forming stability peninsulas are spherical in
DEF HF calculations, 
which is characteristic of magic numbers. The spherical form allows to run
additional check with SPH HF method, where pairing is treated more precisely. Such
additional calculations were done for 
$^{40}$O, $^{74}$S, $^{108}$Fe, $^{166}$Zr and $^{240}$Ba using SkM*, and taking
into account localized states with 
$n \leq 10$ and $l \leq 14 $. The pairing contribution for these nuclei was
equal to zero, which testifies for the magicity of corresponding neutron
numbers. Let us add that $^{74}$S (SkM*, Ska), $^{72}$Si (SkI2) and $^{80}$Ti (SLy4) have
$ N=58$, which is close to $N=56$, whose magicity is discussed in \cite{32}. The treatment of quasi--bound states in the
pairing scheme is done similar to \cite{33,34}. 

The most impressively extended stability peninsula occurs at $N=184$. Fig.~2
shows the fragment of the neutron drip line around $N=126$ and $N=184$. One can
see that that the peninsulas broaden with higher $Z$. The shift of the drip line
compared to HFB calculations 
results from the different  conditions for its determination, namely, zero one neutron separation energy (in the Koopman's approximation) in our case 
and zero chemical potential in the 
HFB method. 

Fig.~3 shows
one neutron separation energies 
for the isotone chain $N = 184$ (SkM* forces). One can see that $S_n$ and
$S_{2n}$ decrease monotonically. The nucleus $^{240}$Ba lies on the edge of stability peninsula, 
its one neutron separation energy is very small $S_n = 0.024$ Mev (DEF HF
calculation) and $S_n = 0.064$ MeV (SPH HF calculation). The last filled
one--particle level 
is 1$j_{15/2}$, which produces the HF potential with the centrifugal
barrier height of 8.58 MeV. 

Two neutron separation energies for isotones corresponding to $N=184$ are shown
in Fig.~3. 
For $Z=56,58,60$ we used the binding energies of the isotopes are not stable
against one neutron emission. This might, of course, 
reduce the precision of $S_{2n}$ values. One can see that DEF HF, SPH HF and HFB
calculations are in good agreement. The nucleus $^{248}$Gd, which has $Z=64$ has
a positive $S_{2n}$ value. Thus for $N=184$ the stability peninsula contains 
isotopes that are stable against both one and two--neutron emission. 
We compared our method to a computer program using a different
numerical approach based on solving the equations on a two-dimensional spatial grid \cite{23.05}, using a delta force for the BCS pairing interactions. 
A typical comparison of the methods can be seen in
Fig. 3.

\begin{figure}[htp]
\includegraphics[totalheight=0.6\textheight]{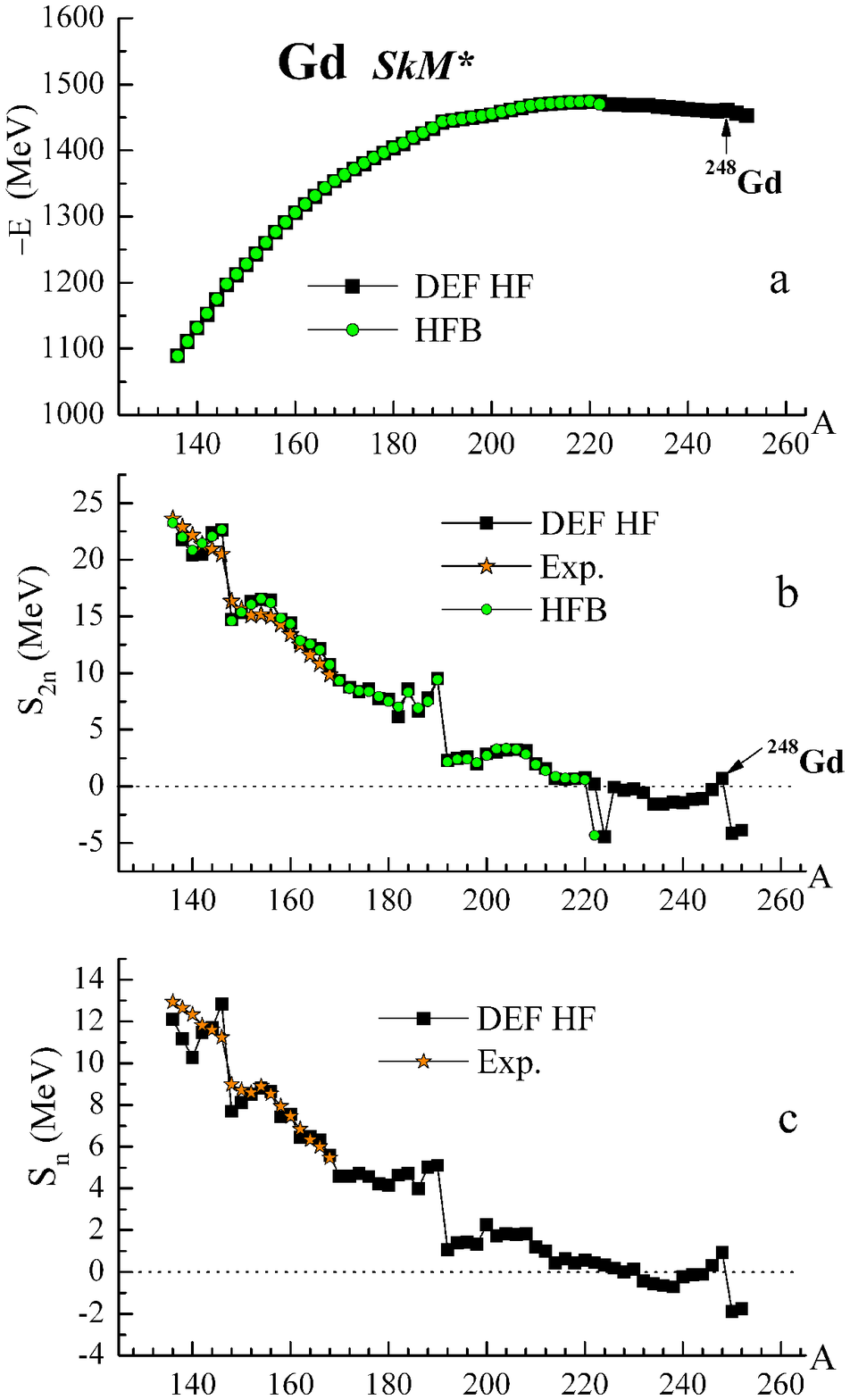}
\caption{(Color online). Total binding energy (a) and two-neutron (b) and one --neutron (c) separation energies  for neutron rich Gd isotopes. Orange stars 
show the experimental data \cite{37}. Dark squares are DEF HF calculations with SkM* forces. Green circles represent the HFB data \cite{4}.}
\end{figure}

In the early papers \cite{35,36} we showed that DEF HF with Skyrme forces provides
a satisfactory description in the region of rare 
earth elements over a wide mass range. Let us take a closer look at the chain of
Gd isotopes. Fig.~4 shows that separation and binding energies are in
good agreement with existing HFB calculations \cite{4}. 
On the figures these values are at some places practically
indistinguishable. The plots excellently fit the existing experimental data
\cite{37} as well. 

In the DEF HF approach the first loss of stability against 2n emission happens at
$A = 222 $, which is close to $A = 220$ obtained in the HFB approach \cite{4}. The
stability is restored for $^{248}$Gd, which is stable both 
against one and two--neutron emission. $^{246}$Gd is stable against one neutron
emission. 
One neutron separation energies, which may be affected by the error of the
Koopmans approximation, agree worse with the 
experimental data than $S_{2n}$ values. One can see the typical bend points at
$N = 82 , 126 , 184$, which come along with magic numbers.

In the Gd isotope chain $^{230}$Gd is the last stable isotope against one neutron
emission, this stability is temporarily lost for higher $A$ and then again
restored for 
$^{246}$Gd. The isotope $^{248}$Gd having the magic number of neutrons $N = 184$ has a spherical shape. 
The isotopes $^{224-230}$Gd are strongly deformed, having $\beta \simeq 0.45$. This
deformation forces the splitting of low--lying quasi stable states with high
angular momentum, and 
some multiplet components enter the discrete spectrum, making $^{230}$Gd stable. 
One can see from Fig.~4(a) that the most stable isotope is $^{222}$Gd and for $^{248}$Gd to decay into $^{222}$Gd it must emit 26 neutrons (the decays into 
the daughter nucleus and 3 or more neutrons have not been observed experimentally so far).

In conclusion, using DEF HF (oscillator basis and grid) and SPH HF approaches with Skyrme forces we show that beyond
the conventional
theoretically predicted 1n drip line there may exist stability peninsulas, which
contain nuclei stable against either 
1n or 2n emission or both. The peninsulas are formed at $N=32, 58, 82, 126,
184$, which are either 
magic or new magic numbers. This was shown for various choices of Skyrme
forces. 
All isotones with such neutron numbers are spherical. The obtained results
indicate that the neutron drip line may have a 
more complicated structure than it was assumed earlier.

 \end{document}